\documentstyle[11pt]{article}
\def\thefootnote{\fnsymbol{footnote}}
\def\bitem{\begin{itemize}}
\def\eitem{\end{itemize}}
\def\be {\begin{eqnarray}}
\def\ee {\end{eqnarray}}
\def\beq {\begin{equation}}
\def\eeq {\end{equation}}
\def\refer {\ref}
\def\ie {{\it i.e.}}

\def\eg {{\it e.g.}}

\def\np {Nucl. Phys.}
\def\prl {Phys. Rev. Lett.}
\def\pr {Phys. Rev.}
\def\pl {Phys. Lett.}

\def\del {\partial}
\def\bi {\begin{itemize}}
\def\ei {\end{itemize}}
\def\ben {\begin{enumerate}}
\def\een {\end{enumerate}}

\def\A{\cal A}
\def\I{\cal I}
\begin{document}
{\it \today}\hfill {\bf NTG-92-20}\\
\vskip 0.4in
\centerline{\large\bf NONABELIAN BERRY PHASES IN BARYONS}
\vskip 0.5in
\centerline{H.K. Lee$^{a}$\footnote{Supported in part by the KOSEF
under Grant No.91-08-00-04 and by Ministry of Education(BSRI-92-231) at
Hanyang University and
through the Center for Theoretical Physics, Seoul National University.},
M.A. Nowak$^{b \ddagger}$\footnote{Supported in part by the
KBN under Grant No PB 2675/2.}, Mannque Rho$^{c}$ and I.
Zahed$^{d}$\footnote{Supported in part by the Department of Energy under Grant
No. DE-FG02-88ER40388 with the State University of New York at Stony Brook.}}
\vskip 0.4in
\centerline{$^{a}$ {\it Department of Physics, Hanyang University}}
\centerline{\it Seoul 133-791, Korea}
\centerline{$^{b}$ {\it Institute of Physics, Jagellonian University}}
\centerline{\it PL-30059, Krakow, Poland}
\centerline{$^{c}$ {\it Service de Physique Th\'{e}orique, C.E. Saclay}}
\centerline{\it F-91191 Gif-sur-Yvette, France}
\centerline{$^{d}$ {\it Department of Physics, State University of New York}}
\centerline{{\it Stony Brook, New York 11794, USA}}

\vskip .5in

\centerline{\bf ABSTRACT}
\vskip .5cm
\noindent We show how generic nonabelian gauge fields can be induced in
baryons  when a hierarchy of fast degrees of freedom is integrated
out. We identify them with nonabelian Berry potentials and discuss their
role in transmuting quantum numbers in bag and soliton models of baryons.
The resulting baryonic spectra both for light and heavy quark systems
are generic and resemble closely the excitation spectrum of diatomic molecules.
The symmetry restoration in the system, {\ie} the electronic rotational
invariance in diatomic molecules, the heavy-quark symmetry in heavy baryons
etc. is interpreted in terms of the vanishing of nonabelian Berry potentials
that otherwise govern the hyperfine splitting.
\par\vfill

\newpage

\renewcommand{\thefootnote}{\arabic{footnote}}
\setcounter{footnote}{0}
\section{Introduction}
\indent

Whenever a quantum system with a hierarchy of length scales is truncated,
induced gauge potentials are naturally
generated reflecting on the degrees of freedom that are
integrated out. A natural
setting for discussing these issues has been discovered by Berry
\cite{berry} in simple quantum systems responding to slowly varying
external parameters. He showed that abelian magnetic monopoles naturally
arise in the  space of the slow variables due to degeneracy points.
This concept has been generalized  by Wilczek and Zee \cite{wz} to the
nonabelian case. They have shown that if a set of degenerate energy levels
depends on adiabatically varying external parameters, nonabelian gauge
potentials are induced, affecting dynamics in a nontrivial way.
Such gauge potentials have attracted a lot of attention in recent years
because of their fundamental character and growing importance in quantum
systems.

The concept of induced gauge fields has led to important understanding
of subtle effects ranging from condensed matter to elementary particle
physics \cite{gp}.
Whenever the underlying dynamics can be separated into ``slow"
and ``fast" degrees of freedom, induced gauge fields are generally expected.
They are generic and embody the essence of the geometrical symmetries
in a given problem. One may therefore ask whether similar gauge structures
are encountered in models of strong interaction physics, and to what extent
they bear on our understanding of hadronic physics.

In a series of recent papers \cite{lnrz}, we have shown that nonabelian Berry
structures can and do appear naturally in topological chiral bags \cite{br}
that model
spontaneously broken chiral symmetry and confinement of QCD
\footnote{While the topological
bag model has confinement and the bag boundary condition plays an essential
role in \cite{lnrz}, we suspect that confinement is not really necessary
for generating gauge structure and that it is only the symmetry that is
relevant.}. The distinction between ``fast" and ``slow" degrees of freedom
is somehow blurred in the topological bag model. However, if we were to
assume that the external pion field can be decomposed into a classical
and quantum part, then a semiclassical delineation is possible in which
the ``slow" degrees of freedom refer to the large component of the fields
and ``fast" degrees of freedom refer to the small component of the field.
In the semiclassical limit the bag is composed of a classical pion field that
wraps valence quarks (bound fermions) and polarizes the Dirac sea. In this
limit neither isospin nor angular momentum are good quantum numbers.
Corrections to this limit are down by $\hbar$ and correspond to
quantum external pions and (multi) quark-antiquark excitations each
of which have good isospin and angular momentum assignment.

In the semiclassical quantization, the classical bag is adiabatically rotated
generating states of good spin and isospin. Even when ignoring the quantum
corrections the adiabatic quantization does not reduce simply to the
quantization of a spinning particle in isospin and spin space as for
ordinary classical fields. Indeed, the degeneracy of the Dirac spectrum,
following the symmetries of the classical field, implies that under
any rotation (even if infinitesimally small) the quarks in the valence orbitals
and the Dirac sea mix inside degenerate bands and between crossing levels.
Adiabatic quark mixing is at the origin of the Berry phases in models
of strong interactions. Below we will analyze their occurrence and
physical relevance in the context of the topological bag model.

This paper is organized as follows. In section 2, we discuss the
general setting for Berry phases in Born-Oppenheimer approximation
and we demonstrate explicitly the Berry phase for the case of diatomic
molecule. We show how the integration of the electronic (fast)
degrees of freedom leads to the extra gauge potential-like term
in the effective Hamiltonian for the nuclear (slow) degrees
of freedom. This term causes splittings of the energy levels
and changes quantum numbers of the system.

In section 3, we show how these concepts extend to the topological
bag model. Subsection 3.1 includes the toy model displaying
all the features of the topological bag model. Subsection 3.2 is
devoted to the bag model itself.
The role of the fast variables is played by the sea quarks
inside the bag, whereas the adiabatic rotation of the solitonic cloud
surrounding the bag constitutes the slow motion. We
detail the explicit Hamiltonian construction
for excited baryons in the light-quark sector
and discuss in subsection 3.3 model-independent mass relations.
In section 4,  we show how the analogous construction of the Berry phases
can be made
in the context of soliton-heavy meson system.
Integrating out heavy meson degrees of freedom we end up with the usual
Skyrme-like rotor term, however, submitted to the influence of a
non-trivial magnetic field (non-abelian Berry phase). This monopole-like
field is responsible for the spin-isospin transmutation of the quantum numbers
and for the structure of the hyperfine splittings. The independent mass
relations are identical to the ones obtained in the framework of the
Callan-Klebanov model of hyperon skyrmions \cite{ck}.

In section 5, we discuss what we believe happens to the skyrmion structure
associated with induced gauge fields
when the heavy meson becomes infinitely heavy at which the recently
discovered heavy-quark symmetry \cite{isgur} is operative.
Our major conclusions and
prospects are relegated to section 6. In Appendix A, we give a heuristic
reason based on an argument by Aharanov et al \cite{aharanov}
why nonabelian Berry potentials cannot vanish in light-quark systems
in contrast to diatomic molecules and heavy-quark baryons.
In Appendix B, an argument is provided as to how heavy mesons decouple
from the Wess-Zumino term responsible for the binding of heavy mesons
to the soliton.

\section{Berry Phases and the Born-Oppenheimer Approximation}
\indent

To define the general setting for Berry phases and help understand
their emergence in the context of the models of elementary particles,
 we will
first present, following \cite{PEDAG,zygel}  the pedagogical example of the
 induced gauge fields (Berry phases)
in the Born-Oppenheimer approximation \cite{moody}.
This approximation is usually described as a separation of slow (nuclear)
and fast (electronic) degrees of freedom. This separation is motivated by the
fact that the rotation of the nuclei does not cause the  transitions
between the electronic levels. In other words, the splittings between the fast
variables are much bigger than the splittings between the slow ones.
We will demonstrate how the integration of the fast degrees of freedom
leads to the induced vector potential of the Dirac monopole affecting the
dynamics of the slow motion. To make our analysis more quantitative, we
define the generic Hamiltonian.
Generically, the Hamiltonian is given by :
\be
H=\frac{\vec{P}^2}{2M} + \frac{\vec{p}^2}{2m} + V(\vec{R},\vec{r})
\ee
where we have reserved the capitals for the slow variables and lower-case
letters
for the fast variables. We expect the electronic levels to be stationary
under the adiabatic (slow) rotation of the nuclei. We split therefore the
Hamiltonian into the fast and slow part,
\be
H&=& \frac{\vec{P}^2}{2M} + h \nonumber \\
h(\vec{R})&=& \frac{\vec{p}^2}{2m} + V(\vec{r},\vec{R})
\ee
where the fast Hamiltonian $h$ depends {\em parametrically}
on the slow variable $\vec{R}$. The snapshot Hamiltonian (for fixed $\vec{R}$)
leads to the Schr\"{o}dinger equation:
\be
h\phi_n(\vec{r},\vec{R}) = \epsilon_n(\vec{R})\phi_n(\vec{r},\vec{R})\,.
\ee
The wave function for the whole system is
\be
\Psi(\vec{r},\vec{R})=\sum_n \Phi_n(\vec{R})\phi_n(\vec{r},\vec{R})\,.
\ee
Substituting the wave function into the full Hamiltonian and using the equation
for the fast variables we get
\be
\sum_n\left[\frac{\vec{P}^2}{2M}
+\epsilon_n(\vec{R})\right]\Phi_n(\vec{R})\phi_n(\vec{r},\vec{R})=
E\sum_n\Phi_n(\vec{R})\phi_n(\vec{r},\vec{R})
\ee
where $E$ is the energy of the whole system.  Note that the operator of the
kinetic energy of the slow variables acts on {\em both} slow and fast
part of the wavefunction. We can now integrate over the fast degrees of
freedom.
A simple algebra leads to the following effective Schr\"{o}dinger equation

\be
\sum_m H_{nm}^{eff} \Phi_m = E \Phi_n
\ee
where the explicit form of the matrix-valued Hamiltonian (with respect to
the fast eigenvectors) is
\be
H_{mn}^{eff} = \frac{1}{2M}\sum_k \vec{\Pi}_{nk}\vec{\Pi}_{km} +
 \epsilon_n\delta_{nm}
\ee
where
\be
\vec{\Pi}_{nm}= \delta_{nm}\vec{P}-
 i<\phi_n(\vec{r},\vec{R})|\vec{\nabla}_r|\phi_m(\vec{r},\vec{R})>\equiv
\delta_{mn}\vec{P}-\vec{A}_{nm}\ .
\ee
The above equation is exact. We see that the fast variables act like a
gauge field. The vector part couples minimally to the momenta, and the
fast eigenvalue acts like the scalar potential.

In the adiabatic approximation one may neglect the off-diagonal transition
terms in the induced gauge potentials, which leads to the simpler
Hamiltonian
\be
H_n^{eff}= \frac{1}{2M}(\vec{P}-\vec{A}_n)^2 + \epsilon_N
\ee
where we denote the diagonal component of the Berry phase (or more precisely
Berry potential) $A_{nn}$ by $A_n$.
If the electronic eigenvalues are degenerate, {\it i.e.}, to the particular
eigenvalue $\epsilon_n$ correspond $G_n$ eigenvectors,
instead of one Berry phase
we obtain the whole set of the $G_n \times G_n$ Berry phases, forming the
matrix
\be
A_n^{k, k^{'}} = i <n, k|\nabla|n, k^{'}>  \,\,\,\,\,\,\,\,k,k^{'}=1,2,...G_n
\,.
\ee
The gauge field so generated is in this case non-abelian and corresponds to the
gauge group $U(G_n)$.
In practical calculations, one truncates the infinite sum in (4) to a few
finite terms. Usually the sum is taken over the degenerate subspace
corresponding to the particular eigenvalue $\epsilon_n$. This is so-called
Born-Huang approximation, which we will use throughout this paper.

Let us finally note that the above formalism may be rewritten in the
Lagrangian language. The corresponding effective Lagrangian is then
equal to\footnote{One can avoid the matrix-valued Lagrangian by using
Grassmannian variables \cite{cabal}. This formalism was used in \cite{lr1}
for describing the same molecular system.}
\be
L_{nm}^{eff}= \frac{1}{2}M \dot{\vec{R}}(t)^2\delta_{mn}
 + i\vec{A}_{mn}[\vec{R}(t)]\cdot
\dot{\vec{R}}(t)-\epsilon_m \delta_{mn}.\label{genlag}
\ee

Let us see how this scenario works for the case of the simple diatomic
molecule. The fast variable describes the motion of the electron around the
internuclear axis.
The slow variables are the vibrations and rotations of the internuclear
axis. This case corresponds to the situation when the energy of the spin-axis
interaction is large compared with the energy splittings between the rotational
levels.  This case is usually called ``Hund case a." We follow the
standard textbook notation of Ref.\cite{LL}.
Let $\vec{N}$ be the unit vector
along the internuclear axis. We can define then the following quantum numbers
\be
\Lambda&=&\,\,{\rm eigenvalue\,\,\,of\,\, }\vec{N}\cdot\vec{L} \nonumber \\
\Sigma &=&\,\,{\rm eigenvalue\,\,\,of\,\, }\vec{N}\cdot\vec{S} \nonumber \\
\Omega &=&\,\,{\rm eigenvalue\,\,\,of\,\, }\vec{N}\cdot\vec{J}=|\Lambda+\Sigma|
\ee
so $\Lambda,\Sigma,\Omega$ are the projections of the orbital momentum,
spin and total angular momentum of the electron on the molecular axis,
respectively.

Let us analyze
the simple case of $\Sigma=0, \Lambda=\Omega=1$.
The fast eigenstates are
\be
|\pm \Omega,\theta,\phi>_S =e^{-i\phi J_3}e^{-i\theta J_2}e^{+i\phi J_3}|\pm
\Omega,0,0>
\ee
where the index $S$ denotes the parametrization singular on the south pole
($\theta=\pi$).
Alternatively, we may use the parametrization
\be
|\pm \Omega,\theta,\phi>_N =e^{-i\phi J_3}e^{-i\theta J_2}e^{-i\phi J_3}|\pm
\Omega,0,0>
\ee
which is singular on the north pole $(\theta=0)$.
The Berry connection is, in our case, a $2 \times 2$ matrix with the following
structure
\be
A_S^{\Omega \Omega^{'}}&=& i _S\!\langle\pm\Omega^{'}, \theta,\phi|d|\pm
\Omega,\theta,\phi\rangle_S \nonumber \\
&=& i _S\!\langle\pm\Omega^{'}, \theta,\phi|\frac{\partial}{\partial \theta}
|\pm \Omega, \theta,\phi\rangle_S d\theta \nonumber \\
& &+ i _S\!\langle\pm\Omega^{'}, \theta,\phi|\frac{\partial}{\partial \phi}
|\pm \Omega, \theta,\phi\rangle_S d\phi\,.
\ee
We can use the orthonormal basis $\vec{A}_S=a_r \vec{e}_r +a_{\theta}
\vec{e}_{\theta}+ a_{\phi}\vec{e}_{\phi}$. A simple calculation
shows that only the $\phi$ component is different from zero, and has
the quasi-abelian form
\be
a_{\phi}^{\Omega,\Omega^{'}} = -\Omega \frac{1-\cos \theta}{\sin \theta}
 (\sigma_3)^{\Omega,\Omega^{'}}
\ee
where $\sigma_3$ denotes the third Pauli matrix, and their components are
numbered by $\Omega,\Omega^{'}=\pm 1$.
An identical calculation based on the parametrization
(14) leads to the expression
\be
a_{\phi}^{\Omega,\Omega^{'}} = +\Omega \frac{1+\cos \theta}{\sin \theta}
 (\sigma_3)^{\Omega,\Omega^{'}}\,.
\ee
We can now calculate the curvature of the Berry connection,
i.e. $F=dA+A \wedge A$. In our simple case the field tensor is quasi-abelian.
We can use any of the gauge fields to calculate the field tensor.
The answer is given by
\be
\vec{B}= {\rm rot} \vec{A} = -\Omega\frac{\vec{N}}{R^3} \sigma_3\,.
\ee
This is nothing else but the magnetic field of the Dirac monopole with the
charge $eg=-\Omega$. We know that the monopole leads to the
observable effects. The kinematical angular momentum operator
gets modified due to the angular momentum stored in the field
of the monopole. A short calculation allows us to extract from the canonical
form (9) the rotational part of the spectrum.
The effective Hamiltonian reads
\be
H^{eff}=\frac{1}{2MR^2}(\vec{J}^2-\Omega^2) + \cdots \label{abelmono}
\ee
where angular momentum operator is given by
\be
\vec{J}= \vec{R}\times \vec{\Pi}-\frac{1}{2}\vec{R} \epsilon_{abc}R_a
F_{bc}=\vec{R}\times \vec{\Pi}-\Omega \vec{N}
\ee
and the ellipsis denotes the vibrational terms.

Of course, the traditional calculation (as one sees in {\it e.g.} \cite{LL}
after correcting a misprint of the factor of 2 in eq. (83.7))
leads to the identical result, modulo
some phenomenological assumptions about the possible spin structure.
We presented the above calculation for two reasons.
Firstly, we believe that this example explains the basic features of
the Berry phase, and while providing a new insight into the
structure of the spectrum of the diatomic molecules, allows us
to understand the modification of the rotator spectrum in terms of simple
physical properties of the Dirac monopole. Secondly, in the following chapters,
we will basically exploit the same strategy to construct a tower of
excited states in the bag model and to make a model-independent analysis
of the soliton-heavy meson bound systems. Since the generic structure
of the obtained spectra for elementary particles is basically similar to
the diatomic mass formulae -- modulo some generalizations due to truly
non-abelian character of the phase which we will shortly sketch-- we
will use the above example as a guide in developing the framework
for describing the technically more complicated systems one encounters in
strong interaction physics.

Before leaving this section, we briefly discuss how the above discussion can
be generalized to a nonabelian situation. The abelian monopole spectrum
corresponds to a special case of diatomic molecule when one restricts the
consideration to the degenerate $\Pi$ doublet with $\pm \Omega$, $\Omega=1$.
For small internuclear distance $R$, the potential
energy curve for the singlet $\Sigma$ with $\Omega=0$ lies higher than that
for the $\Pi$ for which the quasi-abelian approximation is reliable.
However if $R$ is sufficiently large, then the two potential energy curves can
substantially overlap in which case
one must treat the triplets ($\Pi$, $\Sigma$)
together, as pointed out by Zygelman \cite{zygel}. The resulting
Berry potential is then truly nonabelian. The resulting spectrum
can then be written in a {\em generic} form as \cite{lr1}
\be
H^{eff} = \frac{1}{2MR^2} \left(\vec{J}_{\mbox{\tiny R}}+ (1-\kappa)
\vec{J}_g\right)^2  - \frac{1}{2MR^2} (1-\kappa)^2\label{nonabel}
\ee
where
$\vec{J}_{\mbox{\tiny R}}$ is the rotor (``dumb-bell") angular
momentum $\vec{R}\times\vec{\Pi}$ and $\vec{J}_g$ the angular
momentum stored in the nonabelian gauge field, none of which is conserved
separately and the constant $\kappa$ defined by
\be
\kappa=<\Pi|\frac{1}{\sqrt{2}} (L_x-iL_y)|\Sigma>
\ee
where $\vec{L}$ is the electronic orbital angular momentum,
measures how much the rotational symmetry is restored, {\eg}, $\kappa=1$
corresponding to the full restoration of the symmetry. The conserved angular
momentum is $\vec{J}=\vec{J}_{\mbox{\tiny R}}+\vec{J}_g$ which as shown first
by Jackiw \cite{moody} is independent of the charge $[1-\kappa]$.
The limit $\kappa\rightarrow 0$ (small $R$) corresponds to the quasi-abelian
magnetic monopole spectrum (\ref{abelmono}) with  $\Omega=1$.
In the limit
$R\rightarrow \infty$, the singlet $\Sigma$ becomes degenerate with the doublet
$\Pi$ and hence $\kappa\rightarrow 1$. Zygelman shows that in that limit
\be
1-\kappa\sim C/R^4
\ee
where $C$ a constant. In this limit, one can show that the field strength
tensor vanishes (pure gauge). Nonetheless as noted above, there
is an angular momentum associated with the electronic degrees of freedom
which however decouples from the spectrum. What happens is that the electronic
rotational symmetry, broken for small $R$, is restored for large $R$ so that
the electronic angular momentum becomes a good quantum number. This point
will be relevant when discussing the analogy with the heavy quark limit below.

\section{Berry Phase in the Topological Chiral Bag}

\subsection{ Toy Model }
\indent

To fully appreciate the generic structure of the Berry potentials that we
will exhibit, it is useful to reformulate a well-studied case in a way
suitable to our strong-interaction model.
Consider a system of slowly rotating solenoid coupled to a fast
spinning object (call it ``electron") described by the (Euclidean) action
\cite{stone}
\be
S_E=\int dt \left(\frac{{\cal I}}{2}\dot{\vec{n}}^2+\psi^\dagger
(\del_t-\mu\hat{n}\cdot\vec{\sigma})\psi \right) \label{stonem}
\ee
where $n^a(t)$, $a$=1,2,3, is the rotator with $\vec{n}^2=1$, ${\cal I}$
its  moment of inertia, $\psi$
the spinning object (``electron") and $\mu$ a constant. We will assume
that $\mu$ is large so that we can make an adiabatic approximation
in treating the slow-fast degrees of freedom. We wish to calculate the
partition function
\be
Z=\int [d\vec{n}][d\psi][d\psi^\dagger] \delta (\vec{n}^2-1) e^{-S_E}
\ee
by integrating out the fast degree of freedom $\psi$ and $\psi^\dagger$.
This system in the space of the rotating solenoid gives precisely the same
abelian monopole spectrum (\ref{abelmono}) with $\Omega=1/2$.
We will solve this problem first in the standard way used by Stone and then
by the method we shall use.
The procedure used by Stone goes as follows.
Imagine that $\vec{n} (t)$ rotates slowly. At each instant $t=\tau$,
we have an instantaneous Hamiltonian $H(\tau)$ which in our case
is just $-\mu \vec{\sigma}\cdot \hat{n} (\tau)$ and the ``snap-shot"
electron state $|\psi_0 (\tau)\rangle$ satisfying
\be
H(\tau)|\psi^0 (\tau)\rangle=\epsilon (\tau)|\psi^0 (\tau)\rangle.
\ee
In terms of these ``snap-shot" wave functions, the solution of the
time-dependent Schr\"{o}dinger equation
\be
i\del_t |\psi (t)\rangle=H(t)|\psi (t)\rangle \label{sequation}
\ee
is
\be
|\psi (t)\rangle=e^{i\gamma(t)-i\int_0^t \epsilon (t^\prime)dt^\prime}
|\psi^0 (t)\rangle.
\ee
Note that this has, in addition to the usual dynamical phase involving the
energy $\epsilon(t)$, a nontrivial phase $\gamma (t)$ -- known as Berry
phase -- which substituted into (\ref{sequation}) is seen to satisfy
\be
i\frac{d\gamma}{dt}+\langle \psi^0|\frac{d}{dt}\psi^0\rangle=0.
\ee
This allows us to do the fermion path integrals to the leading order
in adiabaticity and to obtain (dropping
the trivial dynamical phase involving $\epsilon$)
\be
Z=const\int [d\vec{n}]\delta (\vec{n}^2-1) e^{-S^{eff}}, \label{z}\\
S^{eff} (\vec{n})=\int \L^{eff}=\int [\frac{{\cal I}}{2}
\dot{\vec{n}}^2-i\vec{\A} (\vec{n})\cdot \dot{\vec{n}}] dt \label{sefff}
\ee
where
\be
i\vec{\A} (\vec{n})=-\langle \psi^0 (\vec{n})|\frac{\del}{\del\vec{n}}
\psi^0 (\vec{n})\rangle \label{bpot}
\ee
in terms of which $\gamma$ is
\be
\gamma=\int \vec{\A}\cdot d\vec{n}.\label{berryphase}
\ee
$\A$ so defined is the Berry potential or connection and $\gamma$
is the Berry phase. $\A$ is a gauge field
with coordinates defined by $\vec{n}$.

We can obtain the same result by defining $S(\tau)$ in
(\ref{stonem}) as
\be
\hat{n} (\tau)\cdot \vec{\sigma}=S(\tau)\sigma_z S^\dagger (\tau).
\ee
We now rotate the electron field as
\be
\psi\rightarrow S\psi.
\ee
Then Eq. (\ref{stonem}) can be written
\be
S_E=S_{S=1} +\int dt \ \left(\psi^\dagger S^\dagger i\del_t S\psi\right).
\label{spluss}
\ee
When the electron field is integrated out, the second term of this action
gives rise to the Berry
potential term of (\ref{sefff}) given in terms of the matrix element
taken with the basis that diagonalizes the fermion term in $S_{S=1}$.
If we call this basis $|\sigma_z>$ ({\ie}, eigenstate of $\sigma_z$), then
\be
i\vec{\A} (\vec{n})=-\langle \sigma_z|S^\dagger\frac{\del}{\del \vec{n}}
S|\sigma_z\rangle.
\ee
This is the procedure that we will use for the more complicated case of the
topological chiral bag.

\subsection{Bag Model}
\indent

The simple description outlined above carries through in spirit to a system of
quarks confined inside a cavity wrapped by a strong pion field. In the bag, the
monopole field is substituted by an induced instanton-like field in isospin
space (the slow variable space) and the heavy fermion is played by valence
quarks. In what follows, we will use the action formulation.

Inside the bag, the quarks are described by free QCD and confined by fiat
at the bag surface. To prevent explicit chiral symmetry breaking a pion
field surrounds the bag. In the topological bag model the pion field has
the structure of the Skyrme hedgehog ansatz.
The latter is invariant under a grand-spin rotation
(angular momentum plus isospin), $\vec{K}=\vec{J}+\vec{I}$.
As a result, the quarks inside the
cavity are polarized in a level structure that depends explicitly on the
strength of the pion field at the bag surface (denoted $F$ and referred
to as ``chiral angle").
 The level degeneracy is $2K+1$. Thorough discussions
on the topological bag model can be found in Ref. \cite{br,chiralbag}.

Suppose that we adiabatically rotate the bag in space. Because of
the degeneracy of the Dirac spectrum, mixing between quark levels
is expected no matter how small the rotation is. This mixing takes
place in each quark $K$-band and leads to a nonabelian
Berry or gauge field. Indeed, an adiabatically rotating bag can
be described by the following action

\be
S_S = \int_V\,\overline{\psi}
i\gamma^{\mu}\partial_{\mu}\psi
-\frac 12\int \Delta_s \,\,
\overline{\psi} \,S\,e^{i\gamma_5\vec{\tau}\cdot\hat{r} F(r)}\,S^{\dag}\,
\psi
+ S_M(SU_0 S^{\dag})\label{gfa1}
\ee
where $V$ is the bag volume (which we shall suppress below
unless ambiguity arises),
$F$ is the chiral angle appearing in $U_0=e^{i\vec{\tau}\cdot\hat{r}
F(r)}$, $\Delta_s$ is a surface delta function and the space rotation
has been traded in for an isospin rotation ($S(t)$)
due to the hedgehog symmetry considered here.
The purely mesonic terms outside the bag are described by $S_M$.
Presently we shall discuss the effect of rotations on quarks only,
relegating the discussion of the mesonic cloud to the second part of this
chapter.  The massless quarks inside the
bag are assumed to be free. The rotation at the boundary can be
unwound by the redefinition $\psi\rightarrow S\psi$, leading to
\be
S_S = S_{S={\bf 1}} +\int\,{\psi}^{\dag}\,{S}^{\dag}i\partial_t S\,\psi\,.
\label{Ss}
\ee
The effect of the rotation on the fermions inside the bag is the
same as a time dependent gauge potential. This is the origin of the induced
Berry potential analogous to the solenoid-electron system , Eq.(\ref{spluss}).

To understand the physics behind this term, we expand
the fermionic fields in the complete set of states $\psi_{KM}$
with energies $\epsilon_{K}$ in the {\it unrotating bag} corresponding to
the action $S_{S=1}$ in (\ref{Ss}), and $M$ labels
$2K+1$ projections of the grand spin $K$. Generically,

\be
\psi (t,x) =\sum_{K,M} c_{KM}(t) \psi_{KM} (x)
\ee
where the $c$'s are Grassmannians, so that

\be
S_S = \sum_{KM} \int dt
\,\, c^{\dag}_{KM} (i\partial_t -\epsilon_K)c_{KM} +
\sum_{KMK^{'}N} \int dt\,\, c^{\dag}_{KM} A^{KK^{'}}_{MN}
                c_{K^{'}N}
\label{action}
\ee
where
\be
A_{MN}^{KK^{'}}= \int_V d^3x\, \psi^{\dag}_{KM} S^{\dag}i\partial_t S
\psi_{K^{'}M}\,.\label{gaugef}
\ee
No approximation has been made up to this point. If the $A$ of (\ref{gaugef})
were defined in the whole $K$ space, then $A$ takes the form of a pure gauge
and the field strength tensor would be identically
zero \footnote{As we saw in the case of the diatomic molecule, the vanishing
of the field tensor does not imply that there is no effect. It describes
the restoration of certain symmetry. See later a similar phenomenon in
heavy-quark baryons.}. However we are forced to truncate the space.
As in the preceding
chapter, we can use now the adiabatic approximation and neglect the
off-diagonal
terms in $K$, {\it i.e.}, ignore the effect of adiabatic rotations, which
can cause the jumps between the energy levels of the fast quarks.
Still, for every $K \neq 0$ the adiabatic rotation mixes $2K+1$ degenerate
levels
corresponding to the particular fast eigenenergy $\epsilon_K$.
  In this form we clearly see that the rotation induces a hierarchy
of Berry potentials in each K-band, on the generic form  identical to
Eq.(10).  This field is truly a gauge field. Indeed,
any local rotation of the $\psi_{KM} \rightarrow D^K_{MN} \psi_{KN}$
where $D^K$ is a $2K+1$ dimensional matrix spanning the representation of
rotation in the $K$-space,
can be compensated by a gauge transformation of the Berry potential
\be
A^K \rightarrow D^K (\partial_t + A^K ) D^{K \dag}
\ee
leaving $S_S$ invariant \cite{jackiw}.

The structure of the Berry potential depends on the choice of the
parametrization of the isorotation $S$ (gauge freedom). For the
parametrization $S =a_4 +i\vec{a}\cdot\vec{\tau}$ with the unitary
constraint $a\cdot a =1$ (unitary gauge), we have
\be
A^K = T_K^a\, A_K^a = T_K^a\,
\left( \,g_K\,\,\frac{\eta_{\mu\nu}^a\,a_{\mu}da_{\nu}}{1+a^2}\right)
\label{berrypot}
\ee
where $\eta$ is the t'Hooft symbol and $g_K$ the induced coupling
to be specified below.  The $T$'s refer to the K-representation
of SU(2), the group of isorotations. In the unitary gauge the Berry
potential has the algebraic structure of a unit size instanton in isospace,
{\ie}, the space of the slow variables. It is not the Yang-Mills instanton,
however, since the above configuration is not self-dual due to the unitarity
gauge constraint.
This configuration is a non-abelian generalization of the monopole-like
solution present in the diatomic molecular case.

To make our analogy more quantitative, let us refer to the
Grassmannians $c$ in the valence states by $\alpha$'s and those in the Dirac
sea by $\beta$'s. Clearly (\refer{action})
can be trivially rewritten in the form
\be
S_S &=& \sum_{KMN} \int dt
\,\, {\alpha}^{\dag}_{KM} \left[(i\partial_t -\epsilon_K) {\bf 1}_{MN} -
                (A^K)_{MN}\,\right] \alpha_{KN} \nonumber\\
     &+& \sum_{KMN} \int dt\,\,
 {\beta}^{\dag}_{KM} \left[(i\partial_t -\epsilon_K) {\bf 1}_{MN} -
                (A^K)_{MN}\,\right] \beta_{KN}\,.
\ee
Integrating over the Dirac sea {\em in the presence of valence
quarks} yields the effective action
\be
S_S =&& \sum_{KMN} \int dt \,\,
 {\alpha}^{\dag}_{KM}\left[ (i\partial_t -\epsilon_K) {\bf 1}_{MN} -
                (A^K)_{MN}\,\right] \alpha_{KN} \nonumber\\
    + && i{\rm Tr}\,{\rm ln}\,
          \left( (i\partial_t -\epsilon_K) {\bf 1}_{MN} -
                (A^K)_{MN}\, \right)
\label{full}
\ee
where the Trace is over the Dirac sea states. The latter can be
Taylor expanded in the isospin velocities $\dot{a}_\mu$ in the
adiabatic limit,
\be
i{\rm Tr}\,{\rm ln}\,((i\partial_t-\epsilon_k){\bf 1}_{MN} - ({\cal
A}_{\mu}^K)_{MN} \dot{a}_{\mu})= \int\, dt \, \frac {{\I}_q}2
\dot{a}_{\mu}\dot{a}_{\mu} + \cdots
\label{sea}
\ee
We have exposed the velocity dependence by rewriting the form
$A^K_{MN}= ({\cal
A}_{\mu}^K)_{MN} \dot{a}_{\mu}$.
Linear terms in the velocity are absent since the Berry phases
in the sea cancel pairwise in the SU(2) isospin case under
consideration. For SU(3) they do not and are at the origin of
the Wess-Zumino term. The ellipsis in (\refer{sea}) refers to higher derivative
terms. ${\I}_q$ is the moment of inertia of the bag. We do not need
the explicit form of this term for our considerations. We would like to
point out that this term includes implicitly the valence quark effect,
 because the
levels of the Dirac sea are modified due to the presence  of the valence
quarks.

To see the general motivation for studying the excited states via Berry phases
let us consider the case of the bag containing one valence quark in the $K=1$
state.
The action for the adiabatic motion of this quark is obtained from the
above formulae and yields
\be
S_S = \int dt\,\,[i{\alpha}^{\dag}_{1M}\dot{\alpha}_{1M}-\epsilon_1
\alpha^{\dag}_{1M}\alpha_{1M} +\frac{1}{2}{\I}_q
\dot{a}_{\mu}\dot{a}_{\mu} + \dot{a}_{\mu}({\cal A}_{\mu}^1)_{MN}
\alpha_{1M}^{\dag}\alpha_{1N}]\,.
\ee
As we will see below, when canonically quantized,
the generic structure of the resulting Hamiltonian is identical to
(\ref{abelmono})
and shows that the excited quark system in the slow variable space behaves as a
spinning charged particle coupled to an instanton-like gauge field
centered in an $S^3$ sphere in the four dimensional isospin space.
This once again illustrates the universal character of the Berry phases.

Let us now quantize the system. Since
$S^3$ is isomorphic to the
group manifold of  $SU(2)$, it is convenient to use the left or right
Maurer-Cartan forms as a basis for the vielbeins (one-form notation understood)
\be
S^{\dag} idS = - \omega_a \tau_a = -v_a^c(\theta) d\theta^c \tau^a\label{s3}
\ee
where we expressed  the ``velocity" forms $\omega$ in the basis  of the
vielbeins $v_a^c$, and $\theta$ denotes some arbitrary parametrization
of the $SU(2)$, {\it e.g.} Euler angles.
In terms of the vielbeins, the induced gauge potential simplifies to
\be
{\cal A}^c= -\, g_K\, v_a^c(\theta) T^a\label{gkd}
\ee
where $T$ are the generators of the Berry potential in the $K$
representation  and $g_K$ is the corresponding charge \cite{lnrz}
\be
g_K = \frac{1}{K}\left( \frac{1}{1+y} \right) -
\frac{1}{K+1}\left(\frac{y}{1+y} \right)\label{berrycharge}
\ee
where
\be
y=\frac{j_{K+1}^2 + j_K^2 -2(K+1)j_{K+1} j_K /x}
{j_{K-1}^2 + j_K^2 -2K j_{K-1} j_K /x}
\cdot  \frac{j_K(1+\frac{\sin F}{2K+1}) - j_{K-1} \cos F}
{j_K(1-\frac{\sin F}{2K+1}) + j_{K+1} \cos F} \nonumber
\ee
and  $j_K$ are the spherical Bessel functions calculated at $x=\omega R$
 -- the  lowest energy solution for fixed $K$ and parity
$P=(-1)^{K+1}$ in a spherical bag. A qualitative behavior of the Dirac
spectrum and the induced charge versus  $F$ are shown in Figs. 1 and 2.
(Note that we could have equally well
used the right-invariant Maurer-Cartan form instead of the left-invariant
Maurer-Cartan form (\ref{s3})). The field strength can be
written  in terms of ${\cal A}$ defined in eq.(\ref{gkd})
\be
{\cal F}_K= d{\cal A}_K-i{\cal A}_K\wedge {\cal A}_K =-g_K(1-g_K/2)\epsilon^{
mij}T_K^m\,v^i\wedge
v^j.
\ee
${\cal F}_K$ vanishes  for $g_K=0$ (trivial case)
and  for $g_K=2$, $i.e.$ the Berry potential becomes a
pure gauge.

The vielbeins -- and hence ${\cal A}$ and ${\cal F}$ -- are frame-dependent,
but
to quantize the system, no specific choice of framing is  needed.
 The canonical momenta are
$p_a= {\partial L}/{\partial \dot{\theta}_a}$.
Our system
lives on $S^3$ and is invariant under $SO(4) \sim SU(2) \times SU(2)$.
Right and left generators are  defined as
\be
R_a &=& u_a^c\,p_c \nonumber \\
L_a &=& D_{ab}(S)\,R_b
\ee
where $u^a_i\,v^i_c = \delta_c^a$ and $D(S)$ spans the adjoint
representation
of the $SU(2)$. Following the procedure described in \cite{lnrz}, we get
our Hamiltonian in terms of the generators\footnote{Canonical quantization
for this system goes much like that of eq.(\ref{genlag}) except that
here we carry along Grassmanians which play an inert role of specifying
the quark states involved, {\i.e.}, equivalent to projection operators.
On the other hand, one can also get eq.(\ref{hstar}) following the
quantization procedure described in
\cite{cabal} for  a system with Grassmanian variables.}
\be
H^* = \epsilon_K \, {\bf 1}     + \frac{1}{8\I}\,
  \left( R_j -g_K\,T_{Kj}\right)
   \left(R_j -g_K\,T_{Kj} \right). \label{hstar}
\ee
This resembles closely the nonabelian molecular Hamiltonian (\ref{nonabel}).
In fact, it is identical to it with a suitable reinterpretation of the
charge $g_K$ to be explained below.
As a result, the Hamiltonian for a singly excited quark\footnote{If we were
to add a second quark to this band (doubly excited state) then we could no
longer have an irreducible representation of $T_{K}$ but a reducible
representation instead.}
takes the simple form
\be
H^* =  \epsilon_K {\bf 1} +\frac{1}{8\I}\left(\vec{R}^2 -2g_K\vec{R}
\cdot\vec{T}_K
        +g_K^2\vec{T}_K^2\right). \label{hstarf}
\ee
The spectrum  can be readily constructed if we notice that (\ref{hstarf})
can be rewritten solely in terms of the independent Casimirs

\be
H^* =  \epsilon_K {\bf 1}+ \frac{1}{2\I}\left[
+\frac{g_K}{2}\vec{J_K}^2 + (1 - \frac{g_K}{2}) \vec{I}^2
-\frac{g_K}{2}(1-\frac{g_K}{2})\vec{T_K}^2 \right]\label{hexcit}
\ee
where $\vec{J}_K =- {\vec{R}}/{2}+\vec{T}_K$ and $\vec{I}=\vec{L}/2$
are the angular momentum and isospin respectively.

The identification of the quantum numbers follows from the original
symmetries of the action. Indeed, under an isospin transformation

\be
S\rightarrow e^{-iT\cdot\alpha} S\qquad\qquad
\psi\rightarrow \psi
\ee
following the redefinition $\psi\rightarrow S^{\dag} \psi$
(isospin co-moving frame). The
isospin operator is given by the standard Noether construction

\be
I^a = D^{ab} (S) \left( {\cal I} \omega^b +
                         \int d^3x \psi^{\dag} T^b \psi \right) =
D^{ab} (S) \left( {\cal I} \omega^b + {g_K} \frac{T^a_K}2 \right).
\label{isospin}
\ee
The second term in (\ref{isospin}) is the induced Berry phase.
The term in bracket is the momentum canonically conjugate to
the velocity $\omega^a$, referred to as $p^a$ above in the
canonical frame. Under a rotation,

\be
S\rightarrow S\,e^{-iT\cdot\beta}\qquad\qquad
\psi\rightarrow \left( e^{iT\cdot\beta}e^{-i J\cdot\beta}\right) \psi.
\ee
Again, the angular momentum is given by the conventional Noether
construction

\be
J^a = -{\cal I} \omega^a + \int d^3x \psi^{\dag} \left( L^a + \frac
{\sigma^a}2\right) \psi =
-\left( {\cal I} \omega^a + {g_K} \frac{T_K^a}2 \right) + \int d^3 x
\psi^{\dag} K^a \psi.
\label{angular}
\ee
Since the states are eigenstates of $K$, the last term in
(\ref{angular}) is just the representation of the SU(2)
algebra spanned by $K$,

\be
J^a = -\left( {\cal I} \omega^a + g_K \frac{T_K^a}2 \right) + T^a_K.
\label{angularx}
\ee
The angular momentum gets an extra contribution due to the induced
non-abelian Berry phase. This is the reason why we are able to
avoid the Skyrme constraint $I=J$ - isospin hidden in the $K$ structure
of the rotated degenerate levels gets transmuted into an extra component
of the angular momentum.

For $g_K=0$, we have the rotor spectrum $H^*={\vec{I}^2}/{2\I}$.
This happens for any value of the chiral angle
only for the $K=0$ level and corresponds
to the known case of the nucleon and delta. For $K>0$, $g_K$ vanishes
for some specific values of the chiral angle (see Fig.2), most probably
connected with the additional level crossings in the spectrum (see Fig.1).
But these may be artifacts and may not be physically meaningful.
For $g_K~=~2$, the Berry field strength
vanishes but the gauge field nontrivially affects the spectrum, {\it i.e.},
$H^*\sim  {\vec{J}}_K^2/2\I$. The spectrum may look
analogous to the quasi-abelian
case of the diatomic molecule but because of the vanishing field strength,
the analogy is not significant.
In our system, however, this situation is never reached as the
charge $g_K$ in Eq. (\ref{berrycharge}) cannot reach 2.
The reason is that there is no limit in which the
adiabatic rotation would be induced by the $K$-spin, and not by the
isospin only. Indeed, if that was the case, we see
immediately from (\ref{angularx}) that the angular momentum of the
system would reduce to the inertial part ${\cal I}\vec\omega$ carried
solely by the hedgehog core. More discussions on this difference
will be given in Appendix A.

There is an amusing analogy with the diatomic molecule above
and the heavy quark system below, if we were to
consider the fictitious situation of two quarks in the $(1^-, 2^-)$,
$(2^+, 3^+)$, $etc.$ states. These multiplets, correspond respectively
to a core with angular momentum $\frac 32^{-}$, $\frac 52^{+}$, $etc.$
coupled to isospin $1/2$. They are the equivalent of the heavy quark
multiplets to be discussed below. As the bag radius is increased
(MIT limit) angular momentum becomes a good quantum number. Thus
the isospin triplet and singlet states become degenerate. In this
limit, the Berry phase stemming from the singlet $exactly$ balances
the Berry phase from the triplet at the MIT point ($F=0$) since
$g_2^- = -g_1^- = 1/2$, $g_3^+ = -g_2^+ =1/3$, $etc.$. This cancellation
does not occur in the lowest multiplet $(0^+, 1^+)$ with an angular
momentum core $\frac 12^+$. The reason is that the Berry phase
vanishes identically in the $K=0$ state for all values of the pion
field $F$. Since they  interpolate between positive and negative energy
levels, these states are not allowed to carry a Berry phase.

To summarize: We see that the role of the induced gauge potential is to lift
the
degeneracy between angular momentum and isospin, and leads naturally
to the description of excited states.
The Hamiltonian (\ref{hexcit}) allows a simple description of the even/odd
parity
excitations of the nucleon and $\Delta$, in terms of the original splittings
in the topological bag model and the induced Berry charge $g_K$. For that
we have to add two quarks in the inactive band $K=0$ each with energy
$\epsilon_0$ and recall that the parity assignment follows from the parity
of the excited quark in the active band $K=1$, which is assumed to describe
the low-lying excited states.

\subsection{Light-quark spectrum}
\indent

A number of  relations among the low-lying excited states of baryons follow
from  (\ref{hexcit}). Here we will only quote some model-independent results
\footnote{In deriving these formulae we have assumed that the
pion cloud outside the bag is not substantially distorted by the excitation of
a single quark inside the bag.}, obtained
by elimination of both the Berry charge $g_1$ and the moment of inertia
${\I}$. For instance, in the Roper channel, it follows from (\ref{hexcit}) that
\be
M(P11)-M(N) =M(P33) -M(\Delta ).\label{ex1}
\ee
Empirically, the left-hand side is $502$ MeV and the right-hand side
is 688 MeV. In the odd-parity channel
\be
M(D13) -M(D35) +M(\Delta ) -M(N) = -\frac 14 (M(D35) -M(S31)).\label{ex2}
\ee
   From the data, we get 116 MeV for the left-hand side and 76 MeV for the
right-hand side.
Also
\be
M(S31)-M(S11) = \frac{5}{2}(M(\Delta) - M(N)) - \frac{3}{2}
(M(D35)-M(D13)).\label{ex3}
\ee
Empirically, the left-hand side gives 85 MeV and the right-hand side gives
125 MeV.

We recall that the above formulae were obtained for the quark sector
only (the interior of the bag), {\it i.e.}, till now we were ignoring
the pionic cloud
{\em outside} the bag, described by $S_M$ in (\ref{gfa1}).
We expect that the detailed
analysis of the pionic sector should give the same structure of the mass
formula. The argument is as follows. Description of the resonances
in the Skyrme model is obtained  by studying phase shifts of the pionic
fluctuations in the background of the static soliton. The adiabatic rotation
(cranking) of the soliton corresponds to slow variables. The pionic
fluctuations are fast and are equivalent to the ``particle-hole" vibrations
in the quark bag. Again, the generic Born-Oppenheimer scenario tells
us that the evolution of the Skyrme cloud outside the bag will be influenced
by the presence of the magnetic force coming from the integrated-out pionic
fluctuations. The counterpart of the charge $g_K$ and moment of inertia  will
of course depend on the version of the Skyrme Lagrangian used, but the generic
formula should be identical.
If we neglect the anharmonicities coming from the vibration
and higher order terms ($O(1/N_C^2)$) coming from the collective rotations
we are at the same level of accuracy in the $1/N_c$ expansion
on both sides of the bag wall, {\it i.e.}, in the quark sector
inside as well as
in the pionic sector outside the bag.  The analysis done recently in
\cite{hawalli} for the pure skyrmion case supports this point
of view.

Pure skyrmion may be viewed as the limiting case of the shrinking bag.
The formulae presented in \cite{hawalli} for the S-wave pion-nucleon
scattering have the same generic form as our mass formula.
We would like to stress that the proper inclusion of the rotational effects
is crucial for the solution of the long standing problem
in the Skyrme like models of the $S11$ and $S31$ degeneracy.
Explicit calculations in \cite{hawalli} (although not relating explicitly
to Berry phases) and our formula (\ref{ex3}) confirm the role of the Berry
phase for splitting the degeneracy between these two levels.
It was noted recently by Masak {\it et al} \cite{masak} that incorporation
of the vector mesons $\rho$ and $\omega$ in a way consistent with hidden
gauge symmetry of chiral Lagrangians \cite{bando}, in particular
in the intrinsic-parity
odd sector, improves markedly the phase shifts for $S11$ and $S31$. The
corresponding processes inside the bag would require additional structure than
what we have been considering and will bring modification to the spectrum,
particularly to (\ref{ex3}).

Finally, let us speculate how bag-radius independent the above formulae are.
In other words, does the Cheshire Cat Principle \cite{cheshire}
(``physics is independent of the bag radius") holds for the excited states?
The structure of the energy levels in the bag
as a function of the skyrmion profile is very complicated.
When changing the bag
radius, several level-crossings are expected to generate additional
contributions to the induced potential.
It can be shown (see Appendix A) using the reasoning of \cite{aharanov}
that modulo a phase
the same field strength tensor can be obtained either from the Berry
potentials constructed within one $K$-subspace -- as in our case --
or from the off-diagonal potentials, connecting different $K$-subspaces.
Therefore, in principle, for a large bag the spacing between the energy
levels becomes increasingly small, so that
some  off-diagonal contributions from {\em e.g.} the $K=0^+,2^+$ levels
crossing  could play an important role.
The point we wish to make is that
the universal character of the Berry phases leaves some hope that
if all the contributions to the gauge potentials are taken into
account on both sides of the bag to  the same order of $N_c$ expansion,
one might expect to obtain an approximate Cheshire Cat picture for the
excited states at the level of the accuracy of the $1/N_c$ expansion.

\section{Berry Phase in Strange Solitons}
\indent

Another interesting application of the above concept is to a system
composed of a soliton and a strange meson.
Strange quarks play a very distinctive role in the strong interaction,
being neither heavy nor light compared with the typical scale of QCD.
A simple but subtle example of the interplay of strange-light degrees
of freedom is provided by the
Callan-Klebanov description \cite{ck} of strange baryons.
In this version of the Skyrme model one assumes {\em ab initio}
that $SU(3)$ flavor symmetry is so badly broken by the massive kaons,
that the usual perturbation theory applied to the mass term in the
Hamiltonian is no longer justified. Kaons are therefore described as
the chiral excitations in the background of the non-strange, $SU(2)$
topological soliton.
The hyperons are then described as molecule-like states
composed of the kaon bound to the soliton. The identification
of the quantum numbers is provided by the usual collective rotations
of the soliton. Adiabatic rotation of the soliton corresponds to the
slow variables, and the kaonic excitations correspond to the fast ones.
We therefore could
expect a Berry phase, which may  influence the dynamics in a non-trivial way.

Here we will describe a simplified model for a system composed of a heavy
meson coupled to a soliton, with an overall isospin invariance. In the
adiabatic limit, the system may be schematically described by
\be
S_A = \int\, dt \left(
-M_H -\frac{{\I}}{4} {\rm Tr}({S}^{\dag}\dot{S})^2 +\int\, d^3x
{K}^{\dag} (t,\vec{x}) \left[i\partial_t + \frac {\nabla^2}{2M_K}-S V(\vec{x})
 {S}^{\dag}\right] K (t,\vec{x}) \right)
\label{meson}
\ee
where ${\I}$ is the moment of inertia of the meson-soliton bound
state, $M_K$ is the meson mass and $V$ is the soliton induced
potential, all of which are model-dependent. Their detailed structure
will not be necessary for our discussion.
We will only mention that the potential distinguishes between the kaons and
anti-kaons in the solitonic background. This is due to the Wess-Zumino term,
which acts as a magnetic like force attracting kaons to the soliton and
repulsing anti-kaons, providing in this way a mechanism for eliminating
spurious
states with $B=1,\,S=1$ from the spectrum. The Wess-Zumino term itself
can be traced back as an abelian Berry phase  coming from the Dirac sea of the
fermionic description  of the original system, but here we would like to
concentrate on the Berry phase coming from the ``heavy" collective
quark-antiquark state ({\em i.e.}\,meson) as described above.

Again, the rotating meson background in (\refer{meson}) can
be unwound through $K\rightarrow S(t)K$ inducing a Berry type term
\be
\int\, dt {K}^{\dag} ({S}^{\dag}i\partial_t S) K\,. \nonumber
\ee
Using the decomposition
\be
K(t,\vec{x}) = \sum_n \, a_n (t) K_n (\vec{x})
\ee
in the {\it unrotated basis}, we can rewrite (\refer{meson}) in the form
\be
S_A = \int\, dt \left(
-M_H -\frac {{\I}}4 {\rm Tr}({S}^{\dag}\dot{S})^2
 +
\sum_{mn}\,  {a_m}^{\dag} \left[ (i\partial_t -\epsilon_m) {\bf 1}_{mn}
            + \int\, dx\, {K_m}^{\dag} ({S}^{\dag}i\partial_t S) K_n\right]
            a_n\right).
\ee
The latter form is totally identical to (\ref{full}) with (\ref{sea})
except that the $a$'s now are c-numbers rather than Grassmannians.
The role of the Berry
potential is to induce hyperfine splitting in the rotor spectrum.
If we denote the eigenenergy of the kaon (or more generally, the
heavy pseudoscalar meson $P=K, D$ as we will discuss later) as $\epsilon$,
then the skyrmion with a bound heavy
$P$ has the fine-structure and hyperfine-structure splitting given
by the Hamiltonian
\be
H=\epsilon +\frac{1}{2\I} \left(\vec{J}_{\mbox{\tiny R}} + c
\vec{T}\right)^2 =\epsilon + \frac 1{2I}\left(\vec J + (c-1)\vec T\right)^2
\label{Heavy}
\ee
where $\vec{J}_{\mbox{\tiny R}}$ is the angular momentum of the rotor
(related to $\vec{R}/2$ of eq.(\ref{hstarf})), $\vec{T}$ the isospin
carried by the meson (or vibration) and $\vec J = \vec J_R +\vec T$
is the total angular momentum of the bound state. $\I$ is the moment of
inertia of the rotor and  $c$ is a constant analogous to the charge
$(1-\kappa)$ in diatomic molecules or to the charge $g_K/2$ of the light-quark
in the chiral bag.
We may immediately write the model-independent formula for this Lagrangian.
It is equivalent to our formulae (\ref{ex1}-\ref{ex3}) and reads
\be
\frac 13\left( 2M(\Sigma^*)+M(\Sigma)\right) - M(\Lambda )=
\frac 23 \left( M(\Delta)-M(N) \right)\,.
\label{CKMASS}
\ee
Experimentally, the left hand side is $304$ MeV and the
right hand side is $293$ Mev.
Originally, this formula was obtained by \cite{ck} without reference to Berry
phases.

\section{Berry Phase in Heavy Solitons}
\indent

Suppose that the strange quark mass becomes so large that it can no longer be
considered as a chiral quark. The question is: As the s-quark mass increases,
say, beyond the chiral symmetry breaking scale, does the concept of skyrmion
with its induced gauge structure still hold?
This is a relevant question since it appears now that the skyrmion picture
holds even when the heavy quark becomes infinitely massive
\cite{manohar,MOPR,NRZ}. The correct
description, however, requires starting {\it ab initio} with a Lagrangian
that satisfies both the chiral symmetry of the light quarks and the Isgur-Wise
(IW) symmetry \cite{isgur,georgi,EFFEC} of the heavy quarks.
The heavy-quark symmetry implies that the pseudoscalar meson $P$ which
plays a key role in the Callan-Klebanov model and the corresponding
vector meson $P^*$ of the quark configuration $Q\bar{q}$ (where $Q$
denotes heavy quark and $q=u,d$ light quark) become degenerate.

Our starting point is the effective action for heavy-light mesons
in the infinite quark mass limit. If we denote by
\be
H= \frac{1+\gamma^0}2 (-\gamma_i P^*_i +i\gamma_5 P)\qquad{\rm and}\qquad
\overline{H} =\gamma^0 H^+\gamma^0
\label{definition}
\ee
the $(0^-, 1^-)$ degenerate doublet in the rest frame of the heavy quark,
then to leading order in the derivative expansion the effective action
follows from \cite{EFFEC,NRZ}
\be
{\cal L}_H= - i{\rm Tr}(\partial_tH\bar{H} )
+{\rm Tr}HV^0\bar{H} -g_H {\rm Tr}H A^i\sigma^i\bar{H} +m_H {\rm Tr} H\bar{H}
\label{mano}
\ee
Here the vector and axial currents are entirely pionic and read
\be
V_\mu = &&+ \frac i2 \left( \xi\partial_\mu\xi^\dagger+
\xi^{\dagger}\partial_\mu\xi \right),  \nonumber\\
A_\mu = && + \frac i2 \left( \xi\partial_\mu\xi^\dagger -
\xi^{\dagger}\partial_\mu\xi\right).
\label{current}
\ee
The pion field $\xi=\exp (i\vec{\tau}\cdot\vec{n}F(r)/2)$
 is described by the usual Skyrme type action.
Alternative formulations involving light vector mesons are also possible
in which case a term of the form
\be
\sim {\rm Tr} \bar{H}H v_\mu B^\mu \label{wzl}
\ee
where $B^\mu$ is the topological baryon current can be generated \cite{NRZ}
and provide a binding mechanism as discussed in \cite{MOPR}. In (\ref{mano})
the parameter $m_H$ is a mass of order $m_Q^0$. For other conventions we
refer to \cite{NRZ}.

The effective action following from (\ref{mano})
is invariant under local $SU(2)_V$ symmetry (h), in which
$V$ transforms as a gauge field, $A$ transforms covariantly and
$H\rightarrow H h^{\dagger}$ and
$\overline{H}\rightarrow h\overline{H}$.
It is also invariant under heavy-quark symmetry $SU(2)_Q$ (S),
$H\rightarrow SH$ and $\overline{H} \rightarrow
\overline{H}S^{\dagger}$. This symmetry mixes the vectors ($1^-$)
with the pseudoscalars ($0^-$). Under the infinitesimal transformation,
\be
\delta{\vec P}^* = \vec\alpha P + (\vec\alpha\times\vec P^*
)\qquad\qquad
\delta P = -\vec\alpha\cdot\vec P^*
\label{Qheavy}
\ee

In the soliton sector the pion field is in the usual hedgehog configuration.
In this case it is useful to organize the $H$ field in K-partial waves.
Generically
\be
H (x,t) = \sum_{KM}  a_{KM}(t)  H_{KM} (x)
\label{Kdecompo}
\ee
where the $a$'s annihilate $H$ particles with good K-spin
where ${\bf K}= {\bf I} +{\bf J} \equiv
 {\bf K}_L + {\bf S}_Q$ with ${\bf I}$ and
${\bf J}$ the total isospin and angular momentum of the H-soliton system,
$K_L$ the K spin of the light antiquark in $H$
and ${\bf S}_Q$ the spin of the heavy quark.
In the original approach of Callan and Klebanov \cite{ck},
 the $K^{\pi}={\frac 12}^{+}$ state
in the kaon channel was found to bind to the soliton.
Can this binding persist in the infinite mass limit?

To answer this question, first let us recall the
essential feature of the Callan-Klebanov scheme which we have
argued above is closely connected to the gauge field hierarchies induced
dynamically. In the Callan-Klebanov scenario the
Wess-Zumino term plays a central role in lifting the degeneracy between the
strangeness $S=\pm 1$ states and assigning the correct quantum numbers to the
physical states. The presence of the
Wess-Zumino term causes P-wave kaons to bind  to the soliton
to order $N_c^0$.
The bound state carries the good grand spin $K=I+J$ (${\frac{1}{2}}^+$)
and heavy-flavor quantum number. However
states with good isospin ($I$) and angular momentum ($J$) emerge after
``cranking" (or rotating) the kaon-soliton bound state as a $whole$.

The origin of the Wess-Zumino term goes back to the underlying fermionic
character of all hadronic excitations.
In Appendix B, we argue that as the mass of $Q$
increases, the topological Wess-Zumino term decouples from the heavy sector
and hence one of the
principal (if not {\it the principal})
agents for the binding needed for a skyrmion with a heavy meson
is lost. The reason can be easily understood. As the mass of the
strange quark is increased (say, to that of the charm or bottom quark),
the Wess-Zumino term truncates to the two-flavor sector which is
identically zero. This is because the heavy mesons can no
longer be viewed as {\it angle} excitations of the chiral order parameter
in the QCD vacuum. The rate at which the Wess-Zumino term disappears
depends on detailed dynamics. Our qualitative arguments suggest that
the rate is controlled by the ratio of the induced constituent masses
$\sigma$. For strange quarks this ratio is : $\sigma /\sigma_S \sim 0.47$
while for bottom quarks this ratio is : $\sigma /\sigma_B \sim 0.32$.

What is the fate of the heavy-meson-skyrmion bound state
when the Wess-Zumino term vanishes?  Two mechanisms
providing classical binding were proposed \cite{NRZ,MOPR}. In the first
approach the binding depends on the form of background pionic potential,
in the second it is strengthened by the vector-meson induced term (\ref{wzl}).
If the binding persists even in the heavy quark
limit then our previous discussion carries through nicely. Indeed as
the mass of the heavy quark is raised, the Berry phase receives
contribution from both the $P$ and $P^*$. Generically

\be
H=\epsilon +\frac{1}{2\I} \left(\vec{J}_{\mbox{\tiny R}} + c
\vec{T} +c_* \vec{T}_*\right)^2
\label{heavy1}
\ee
where $\vec T_*$ is the
isospin contribution of $P^*$ to the
induced Berry phase.
More explicitly, this formula can be rewritten
as \cite{NRZII}
\be
H = \epsilon + \frac 1{2\cal{I}}\left(
(\vec{\bf J} -\vec{\bf S}_H) - (1-C_P){\rm Tr}(P\vec{\bf I}P^+) -
(1-C^*_P){\rm Tr}(P^*_j\vec{\bf I}P_j^{*+})\right)^2
\label{modifiedck}
\ee
where $I,J$ are isospin and angular momentum operators, respectively,
and $S_H$ is the total spin of the H-particle.
In the $K=\frac 12^+$
shell it reduces to $\vec{\sigma}/2$, {\it i.e.} spin $1/2$
representation. This shows that the spin of
the heavy quark and the light quark $fractionate$ in the K-representation.
The H particle bound
to the skyrmion resembles a heavy fermion with spin $1/2$. A similar
transmutation occurs in the Callan-Klebanov construction \cite{ck}. This
fermionization of the original bosonic degrees of freedom through the
hedgehog structure is what makes skyrmions so remarkable.
This result carries to higher K-shells.

In general $C_P$ and $C_P^*$ are complicated functions
of the heavy quark mass. However, in the heavy quark limit $C_P= -C_P^* =1$
and one recovers the rotor spectrum. This cancellation is guaranteed by
two facts : heavy quark symmetry that implies the same strength for $C_P$
and $C_P^*$ and the underlying hedgehog character of the skyrmion that
forces the isospin in $P^*$ to be antiparallel to the spin, flipping
the sign of $C_P$ compared to $C_P^*$.
In the infinitely heavy quark
limit Hamiltonian for the $K=\frac{1}{2}^+$ shell takes the form
\cite{MOPR,NRZII}
(to order $m_Q^0N_c^{-1}$)
\be
H^{\frac 12} =
\frac {{\bf J}^2_R}{2\cal{I}}
 = \frac {(\vec{\bf J} - \vec{\bf S}_H)^2}{2\cal{I}}
 = \frac {{\bf I}^2}{2\cal{I}}\,.
\label{hlight}
\ee
Thus to order
$m_Q^0N_c^{-1}$ the $\Sigma$ and $\Sigma^*$ are degenerate.
Note that the situation is totally analogous to the non-abelian
molecular case for $R\rightarrow \infty$ discussed in Section 2.

 The above Hamiltonian implies the following
mass relation in the heavy hyperon spectrum
\be
\left( M(\Sigma^*) - M(\Lambda )\right)=
\frac 23 \left( M(\Delta)-M(N) \right)
\label{ckmassheavy}
\ee
If we were to ignore  $P^*$ for any finite $m_Q$ then (\ref{modifiedck})
reduces to
\be
H_1 = \frac 1{2\Omega}\left(
\vec{\bf J}  - (1-C_P){\rm Tr}(P\vec{\bf I}P^+) \right)^2.
\label{ckhamiltonian}
\ee
The latter reduces to (the incorrect) $H = {\bf J}^2/2\Omega$ as opposed to
(the correct) $H = {\bf I}^2/2\Omega$ in the heavy quark limit.\footnote{
The heavy-meson limit of the Callan-Klebanov model with the Skyrme quartic
term or with vector mesons as studied in \cite{RRS,omrs} do not go to
$H={\bf J}^2/2\Omega$ since part of the $P^*$ contribution is included
in the treatment. It does not go to the correct heavy limit either.}

It is interesting to ask which of the mass formulae,
 (\ref{CKMASS}) or (\ref{ckmassheavy}),
  works better for the charm sector. The direct comparison is impossible
at the moment, since the mass of the $\Sigma^*$ is not yet measured.
The mass spectrum predicted according to the Callan-Klebanov scheme
-- and generalized for more than one heavy mesons -- for the
charm baryons is given in \cite{omrs}. There the $\Xi$'s and $\Omega$'s
are described by binding the $K$'s and $D$'s without interactions, that is to
say, in quasiparticle approximation. The prediction of \cite{omrs} which does
not manifestly respect the Isgur-Wise symmetry is nonetheless surprisingly
close
to that of quark models, suggesting that perhaps the mass of the charm
quark is not large enough to see clearly the effect of the Isgur-Wise symmetry
at the level of mass formulae. The {\it effective} hyperfine coefficient
$c$ as defined in (\ref{Heavy}) comes out to be 0.62 for the strange hyperons
and 0.14 for the charmed hyperons. The latter is small, but certainly not
near zero as would be the case if the charm quark were massive enough to
satisfy the Isgur-Wise symmetry.

Let us finally note that an approach to the heavy
solitons similar in spirit to what was discussed above
was suggested recently by Manohar and collaborators \cite{manohar}.
The difference is that we have insisted on the mechanism of binding
at the classical level (in \cite{manohar} binding has quantum mechanical
nature) and we rely on the concept of the Berry phases.
 In our approach the $P$ and $P^*$ are defined in the (isospin)
co-moving frame making their quantization simpler for the  bound state
problem since they do not carry good isospin (they carry good K-spin).
The $dressed$ $P$ and $P^*$ used by Manohar and collaborators are defined
in the laboratory frame and their quantization is simpler for the scattering
problem since they carry good isospin (as asymptotic $P$ and $P^*$ do).
The two descriptions are related by a global isospin rotation.
Since both descriptions have built in  heavy quark
symmetry, they yield similar physical predictions. Indeed it is not
difficult to also formulate Manohar's approach in such a way that the
Isgur-Wise
symmetry is realized as the vanishing of the Berry potential defined in the
laboratory frame \cite{lr2}.

\section{Conclusions}
\indent

The topological bag model offers a suitable setting for discussing
Berry phases. The analogy with the fermion-monopole system is striking.
In the bag, the strong pion field distorts the Dirac spectrum causing
the emergence of Berry phases under any adiabatic rotation. While the
Dirac sea produces no net Berry (Wess-Zumino) contribution due to
pairwise cancellations in the sea, the valence states do. The net effect
is similar to a spinning charged particle coupled to an instanton-like
gauge field in isospin space.

The role of the Berry phase is to induce hyperfine splitting in the rotor
spectrum. This effect can be used to describe excited baryons
in the light-quark nonstrange sector. The model-independent relations discussed
here are in fairly good agreement with the data.
Given the simplicity of the description this is striking.

We have argued that the features displayed in the context of the topological
bag model are in fact generic. They can be easily extended to strange
baryons as discussed by Callan and Klebanov and even to heavier systems
as the ones discussed  by Manohar and collaborators
\cite{manohar} and also by Min and collaborators \cite{MOPR}.
This is hardly a surprise given the generic character of
Berry phases.

Our work is certainly far from complete. We have not investigated
systematically  the relevance of the Cheshire Cat description for the
excited states, nor have we explored totally the heavy light systems.
Moreover, we should be also able to address exotic issues related to
photoproduction mechanisms and dibaryon systems where excited quarks
are naturally triggered. We hope,  however,  that our initiative will
spur more excitement  in these directions.

\subsection*{Acknowledgments}
One of us (MR) is grateful for useful discussions with Y. Aharonov, in
particular for explaining his approach of Ref.\cite{aharanov}.
This work has been supported in part by a DOE grant DE-FG02-88ER40388 and
by KBN grant PB 2675/2.

\newpage
\subsection*{Appendix A: The Nonvanishing of Nonabelian Berry Potentials
in the Chiral Bag}
\renewcommand{\theequation}{A.\arabic{equation}}
\setcounter{equation}{0}
\indent

In this Appendix, we wish to explain the difference in structure
between the induced field in light-quark systems and the one in heavy-quark
systems and also in diatomic molecules. We noted in the main text that
while the structure of Berry potentials and their physical effects are
generic,  the field tensor behaved differently. To be specific,
in heavy baryons and diatomic molecules, both the Berry potential and
its field strength vanished in certain limit while they did not in the
chiral bag modeling the light-quark baryons \footnote{It is possible to
construct a chiral bag that includes heavy mesons for which case one should
also have a vanishing Berry potential in heavy-meson limit. See \cite{pmr}
for a recent discussion on this.}.

We first note that when the condition of adiabaticity is satisfied,
there are two ways of describing Berry potentials. One is the standard way
used by Berry \cite{berry} which is to define the potential within
a diagonal subspace (denoted ${\cal A}$) and the other proposed by
Aharanov et al. \cite{aharanov} is to define it in off-diagonal subspaces
(denoted $\tilde{\cal A}$). To define these quantities precisely, we use
the Hamiltonian formalism  of Ref.\cite{giler}. Let the {\it fast variable}
Hamiltonian parametrized by $ a_{\mu}$ at a given time t be written in the form
\be
H(a_{\mu})=\sum_K \epsilon_K (a_{\mu}) \Pi_K (a_{\mu})
\ee
where
$\Pi_K (a_{\mu})$ is the projection operator onto the subspace (labeled by
the index $K$) spanned by the `snap-shot'
energy eigenstate of $\epsilon_K (a_{\mu})$,
\be
H(a_{\mu}) |K, a_{\mu}\rangle =\epsilon_K (a_{\mu})
|K, a_{\mu}\rangle.
\ee
The quark action in eq.(\ref{gfa1}) discussed in ref.\cite{lnrz} can clearly be
quantized to take this generic form. In {\it adiabatic approximation}
the standard form of the Berry potential that is inherited from the {\it fast}
space can be written as
\be
{\cal A}=\sum_K \Pi_K S^\dagger dS \Pi_K \label{berrya}
\ee
where the dependence of the projection operator on the coordinates $a_\mu$
is suppressed and
\be
|K, a_\mu (t)\rangle=S (t) |K, a_\mu (0)\rangle.
\ee

Let us now define formally the off-diagonal field, $ \tilde{\cal A}$, that
connects different subspaces
\be
\tilde{\cal A}=\sum_{K\neq K'} \Pi_K  S^\dagger dS\Pi_{K'} .
\ee
This is the gauge potential of Aharanov et al. \cite{aharanov} up to
a unitary transformation. Calculating the field strength with $\tilde{\cal A}$
we get
\be
{\cal F}_{\tilde{\cal A}}
 =\sum_{K^"} \Pi_{K^"} \left( \tilde{\cal A}\wedge \tilde{\cal A}
\right) \Pi_{K^"}. \label{tident}
\ee
Thus as discussed in Ref.\cite{aharanov,stone2},
although we have used the field $\tilde{\cal A}$ that only mixes
different spaces, the field tensor is diagonal.
Now let us calculate the field tensor with the diagonal field
${\cal A}$ (\ref{berrya}). Using the properties of the projection operator,
one can readily verify that one obtains exactly the same expression as
(\ref{tident}) except for a minus sign
\be
{\cal F}_{\cal A} =  - {\cal F}_{\tilde{\cal A}}.\label{faat}
\ee
This is of course a direct consequence of the fact that ${\cal F}$
originates from the diagonal of $S^{\dag} dS$, and the latter is
a pure gauge in the full Hilbert space.

The way this relation might impact on our discussion is as follows.
In subsections (3.2) and (3.3) we focused on the $K=1^+$ band which is
the first excited $K$ band above the ground band $K=0^+$. However, the $K=1$
band can be connected by the adiabatic rotation operator $S$ to not only
the $K=0$ band but also to the $K=2$ band. As the bag radius increases
or equivalently the chiral angle $F$ tends to zero, the $K=0,1$ bands cross
each other (following the restoration of the angular momentum into the
Dirac spectrum). The $K=2$ band never crosses any of them at any point of
the chiral angle since it carries different angular momentum.
This should be contrasted with the molecular case or
with the heavy-baryon case. In the diatomic molecule, the doubly degenerate
$\Pi$ states cross the singlet $\Sigma$ at $R=\infty$ at which point
the rotational symmetry is restored in $L=1$. What (\ref{faat}) says is that
sufficiently far away from the triple degeneracy point, one can describe
the spectrum either with the diagonal field or with the off-diagonal field.
An analogous situation holds for the heavy-baryon case where the singlet
$P$ ``crosses" the triplet $P^*$ in the IW limit. Thus what is different in
the chiral bag case is that there is {\it no point} at which all the relevant
$K$ states, namely $K=0,1,2$, become degenerate.

To make the above statements more quantitative,
Let us make an {\it ansatz} for a Berry potential that captures the essence
of the above structure. We take in $K$ space
\be
{\bf A}_{KK'} = {\cal A} \, \delta_{KK'} + \rho_{KK'} \tilde{\cal A}_{KK'}
\label{mberry}
\ee
where we have introduced the ``suppression factor" $\rho_{KK'}$ for
$K \neq K'$ for which we make the simplest possible assumption,
\be
\rho_{KK'} &=& 1, \,\,\,\,\,{\rm for} \,\,\, |\epsilon_{K}-\epsilon_{K'}|
\ll \Delta,\label{rho1}\nonumber\\
           &=& 0, \,\,\,\,\,{\rm otherwise}.\label{rho0}
\ee
Here the $\Delta$ represents the scale of the adiabaticity of the slow-variable
system.  The standard Berry potential is recovered when the adiabatic
change of state ({\ie}, the complete suppression of off-diagonal transitions)
is applicable, that is to say, $\rho=0$. Now we calculate
the field strength of ${\bf A}$ using eq.(\ref {mberry}),
\be
{\cal F}_{\bf A}^K &=& \Pi_{K} {\bf A} \wedge {\bf A} \Pi_{K}\nonumber\\
&=& \sum_{K' \neq K} \left( 1 \, - \, |\rho_{KK'}|^{2} \right)
\tilde{\cal A}_{KK'} \wedge \tilde{\cal A}_{K'K}\label{frho}
\ee
where the superscript $K$ on the field strength means that we are focusing
on a particular $K$ space. Although it is obtained with a specific
{\it ansatz}, we believe (\ref{frho}) to be generic.
To see that it is quite general, consider the diatomic molecular case
\cite{zygel,lr1}. As the internuclear distance $R$ becomes large,
the energies of the $\Pi$ and $\Sigma$ levels become degenerate and
hence $\rho_{\Sigma \Pi}=1$.  Therefore
\be
{\cal F}^{\Sigma} = 0 = {\cal F}^{\Pi},\label{fzyg}
\ee
implying the vanishing of the induced interaction\footnote{In fact, if the
$\Pi$ and $\Sigma$ levels are degenerate, then the induced gauge potential
is really a pure gauge which can be gauged away so that ${\bf A} =0$ and
${\cal F}_{\bf A} =0$.}.

Let us now use eq.(\ref{frho}) to show
that in contrast to the diatomic molecule,
there is no such limit in the chiral bag for light-quark baryons for either
the gauge field or the field tensor to vanish. Suppose such a limit existed
in the topological bag model. Then from the above discussion, we should expect
{\it all} the relevant energy levels connected to the reference $K$ level
by the adiabatic rotation operator $S$ to become degenerate.
But in the chiral bag model with the charges given by (\ref{berrycharge}),
this cannot happen. What happens is that when the chiral
angle $F(R)$ goes to $0$, the $K=0,1$ levels become degenerate. So from
eq.(\ref{frho}), for $K=1$,
\be
{\cal F}_{\bf A}^{K=1} &=& \left( 1  -  |\rho_{10}|^{2} \right)
\tilde{\cal A}_{10} \wedge \tilde{\cal A}_{01}+
\left( 1  -  |\rho_{12}|^{2} \right) \tilde{\cal A}_{12} \wedge
\tilde{\cal A}_{21},\label{frho0}\\
 &=& \left( 1 \, - \, |\rho_{12}|^{2} \right)
\tilde{\cal A}_{12} \wedge \tilde{\cal A}_{21} \label{frho1}
\ee
since $\rho_{10}=1$ from our {\it ansatz} (\ref{rho0}).
However $\rho_{12} = 0$ since the $K=2$ state is
still split from $K=0,1$ states.  Therefore (\ref{frho1}) need not vanish.
A similar observation can be made as the bag radius goes to zero, although the
nature of level crossings  is somewhat different.

In the derivation of the Hamiltonian for the excited states,
eq.(\ref{hexcit}), the adiabatic approximation has been assumed to be valid.
This has led to
the bag-radius-independent (``Cheshire Cat") mass relations among the excited
baryons as discussed in the text.  The repeated level crossings, however,
may invalidate some of these approximations. Also, as the bag radius is
increased the level spacing decreases suggesting also the breakdown of the
adiabatic approximation.
This seems to suggest that the
so-called ``bag-radius-independent" mass relations cannot hold and hence
the Cheshire Cat Principle must be breaking down for the excited states.
Nonetheless the mass formulas in subsection 3.3 worked fairly well.
How do we understand this?

The answer may lie in the fact that there is {\it no limit} at which the
field tensor (or the gauge potential) vanishes.
At the bag radius at which the adiabaticity condition presumably fails to
hold, the off-diagonal contributions could significantly modify the charge
$g_1$ from the value implied by (\ref{berrycharge})
in a way suggested by eq.(\ref{frho0}). However since $\left(g_{K}\right)_{R}$
cannot 2, the structure of eq.(\ref{hexcit}) from which the
same mass relations follow will remain unmodified.

\newpage
\subsection*{Appendix B: The Vanishing of the Wess-Zumino Term for
Heavy Quark}
\renewcommand{\theequation}{A.\arabic{equation}}
\setcounter{equation}{0}
\indent

Consider QCD with two massless quarks and a heavy quark of mass $m_Q$.
Generically, the effective action in the single gluon-exchange approximation
to QCD can be rewritten as follows (using Euclidean conventions)
\be
S [ S, P ] = -N_C {\rm Tr\, Ln } \left( \rlap/\partial + m + S +iP\right)
\label{seff}
\ee
where $S$ and $P$ are scalar and pseudoscalar $3\times 3$ hermitian matrices
in flavor
space and $m$ is short for $m= m_Q ({\bf 1} -\sqrt{3} \lambda_8 )/3$. A
comprehensive discussion of (\ref{seff}) can be found in Ref.\cite{ball}.
Without loss of generality, we can use the decomposition
\be
S+iP = \Sigma \, \, e^{i\gamma_5 \phi^a T^a} \equiv \Sigma\,\,U^{\dag}_5
\ee
Standard arguments show that the $\phi$'s could be interpreted as pseudoscalar
mesons and that $\Sigma$ can be related to the dynamically generated
(or ``constituent") quark mass in the
vacuum \cite{ball}. Since the argument in (\ref{seff}) is nonhermitian
operator,
the effective action develops both a real ($S_R$) and imaginary part ($S_I$).
The latter follows from
\be
\frac{\delta S_I}{\delta \phi^a} = \frac 12 \left( (\rlap/\partial + m
+  \Sigma\,\,U^{\dag}_5)^{-1}
       \Sigma\frac{\delta U^{\dag}_5}{\delta\phi^a} - {\rm h.c.}\right)
\label{wzw}
\ee
and gives rise ``usually" to the Wess-Zumino term. If the mass of the heavy
quark becomes large  the Wess-Zumino term vanishes
in the three flavor case.

Indeed, let us assume that the constituent masses are triggered by the
quark condensation in the vacuum. The details by which this occurs is
certainly model-dependent, however, the generic trend is not. Generically
the quark condensate is given by
\be
<\overline{\Psi} \Psi > = -i\int d\lambda \frac 1{\lambda +im} \rho (\lambda )
\ee
where $\rho (\lambda )$ is the distribution of the eigenvalues of the Dirac
operator in Euclidean space. For massless quarks it reduces to
\be
<\overline{q} q > =-\pi {\rm sgn \,\,m }  \,\, \rho (0),
\ee
whereas for heavy quarks $m_Q >>\kappa $ -- the half width of $\rho (\lambda)$,
 typically of  the order of $\Lambda$ in QCD --, we have
\be
<\overline{Q} Q > =-\frac 1{m_Q} \int d\lambda \rho (\lambda )\,.
\ee
Typically, the eigenvalues have a Gaussian distribution (following the
randomness prevailing in the QCD vacuum) so that
\be
\rho (\lambda ) \sim \rho (0)\ \exp \frac{-\lambda^2}{4\kappa^2}
\ee
leading to
\be
\frac {<\overline{Q} Q > }{<\overline{q} q > } =
\sqrt{\frac 2{\pi}}\frac{\kappa}{m_Q}
\label{cond}
\ee
This shows that the heavy quark condensate in the vacuum vanishes as $1/m_Q$.
Up to this point, the arguments are general.

To be able to relate (\ref{cond}) to the ``constituent" masses,
$\Sigma ={\rm diag} ( \sigma, \sigma, \sigma_Q )$, we need to resort to a model
description of the vacuum. Sum rule arguments combined with the constituent
quark model \cite{IOFFE} suggest that
\be
\frac{\sigma_Q}{\sigma} =
\left( \frac {<\overline{Q} Q > }{<\overline{q} q > }\right)^{1/3} =
\left( \sqrt{\frac 2{\pi}}\frac{\kappa}{m_Q}\right)^{1/3}\label{SUM}
\ee
which shows that the ``constituent" mass vanishes as the inverse cubic root
of the heavy current quark mass in the limit where $m_Q >> \kappa \sim
\Lambda$. It is possible that other models lead to a somewhat different
scaling, but we believe the estimate (\ref{SUM}) is good enough to gain
some idea how things might go. In the heavy quark limit $\sigma_Q \sim 0$.

With the above in mind, we can evaluate the Wess-Zumino term through the
standard derivative expansion, using for the propagator
\be
\left( \rlap/\partial + m +\Sigma \right)^{-1} =
\frac{\rlap/\partial -\sigma }{\partial^2 - \sigma^2}{\bf 1}_2 +
\frac{\rlap/\partial - m_Q }{\partial^2 - m_Q}{\bf 1}_3
\ee
where ${\bf 1}_2 = {\rm diag} (1,1,0)$ and ${\bf 1}_3 = {\rm diag} (0,0,1)$.
Since ${\bf 1}_2\,\,{\bf 1}_3 =0$ it is straightforward to show that the
heavy quark contribution drops from the Wess-Zumino term, and one is left
only with the two-flavor (chiral quark) Wess-Zumino term
that is known to vanish.

To summarize, we have shown that as the quark mass $m_Q$ becomes considerably
larger than $\kappa\sim \Lambda$ -- the width of the eigenvalues distributions
of the Dirac operator in the vacuum --, the heavy quark decouples and the
Wess-Zumino term truncates to the two-flavor Wess-Zumino term which is
identically zero. Our reasoning sketched above could presumably
allow one to make an explicit calculation of the rate at which the
Wess-Zumino term vanishes with the heavy-quark mass.


\newpage
\centerline{\bf FIGURE CAPTIONS}
\vskip 1cm
\noindent {\bf Figure 1}
\begin{quotation}
\noindent
Schematic quark spectrum $\epsilon_K R$ (where the subscript $K$ stands for the
grand spin of the quark level) in the chiral bag wrapped by hedgehog pions
as function of the chiral angle $F(R)$. Note that $F(0)=-\pi$.
For a realistic spectrum, see Mulders in Ref.\cite{chiralbag}.
\end{quotation}
\vskip 0.5cm
\noindent {\bf Figure 2}
\begin{quotation}
\noindent
Schematic plot of the ``Berry charge" $g_K$ where $K$ is the grand
spin of the quark level as function of the chiral angle $F(R)$. For a
more realistic plot, see Ref.\cite{lnrz}.
\end{quotation}

\end{document}